\documentclass[11pt]{article}
\usepackage{geometry,latexsym,amssymb,amsfonts,amsmath,xspace,url}                
\usepackage{graphicx}
\usepackage{epstopdf}
\usepackage{amsthm}

\newif\iftodo
\todotrue    

\newtheorem{theorem}{Theorem}

\newtheorem{lemma}[theorem]{Lemma}
\newtheorem{corollary}[theorem]{Corollary}
\newtheorem{definition}[theorem]{Definition}

\newcommand{\tuple}[1]{\langle #1 \rangle}

\usepackage{xcolor}

\newcommand{\notf}{\ominus}
\newcommand{\orf}{\oplus}
\newcommand{\andf}{\otimes}
\newcommand{\impf}{\Rightarrow}

\newcommand{\ie}{\textit{i.e.},\xspace}
\newcommand{\eg}{\textit{e.g.},\xspace}

\newcommand{\nd}{\noindent}
\newcommand{\alc}{${\cal ALC}$}
\newcommand{\bottomc}{\perp}
\newcommand{\topc}{\top}

\newcommand{\impc}{\sqsubseteq}

\newcommand{\defc}{\equiv}
\newcommand{\cass}[2]{\mbox{$#1$:$#2$}}
\newcommand{\rass}[3]{\mbox{$(#1,#2)$:$#3$}}
\newcommand{\andc}{\sqcap}
\newcommand{\all}{\forall}
\newcommand{\some}{\exists}
\newcommand{\notc}{\neg}

\newcommand{\rimp}{\rightarrow}
\newcommand{\requiv}{\leftrightarrow}
\newcommand{\orc}{\sqcup}
\newcommand{\csome}{\exists}

\newcommand{\fuzzyg}[2]{\mbox{$\tuple{#1,#2}$}}

\newcommand{\I}{\mathcal{I}}
\newcommand{\T}{{\cal T}}
\newcommand{\A}{{\cal A}}
\newcommand{\K}{{\cal K}}
\newcommand{\KB}{{\K}}

\newcommand{\highi}[1]{{#1}^\I}

\newcommand{\unit}{[0,1]}

\title{On the Undecidability of Fuzzy Description Logics with GCIs with \L ukasiewicz $t$-norm }
\author{
\begin{tabular}[t]{cc}
    Marco Cerami & Umberto Straccia \\
 \small{IIIA - CSIC} & \small{ISTI - CNR} \\
  \small{Bellaterra, Catalunya} & \small{Pisa, Italy} \\
   \small{cerami@iiia.csic.es } & \small{umberto.straccia@isti.cnr.it} \\
\end{tabular}
  }
  
\date{\today}                                           

\begin{document}
\maketitle

\begin{abstract}
Recently there have been some unexpected results concerning Fuzzy Description Logics (FDLs) with General Concept Inclusions (GCIs). They show that, unlike the classical case,  the DL \alc~with GCIs does not have the finite model property under \L ukasiewicz Logic or Product Logic and, specifically, knowledge base satisfiability is an undecidable problem for Product Logic. We complete here the analysis by showing that  knowledge base satisfiability is also an undecidable problem for \L ukasiewicz Logic.
\end{abstract}

\section{Introduction}

\nd {\em Description Logics} (DLs)~\cite{Baader03a} play a key role
in the design of {\em Ontologies}. Indeed, DLs are important as
they are essentially the theoretical counterpart of the \emph{Web
Ontology Language OWL 2} \cite{OWL2}, the standard language to
represent ontologies.

It is very natural to extend DLs to the fuzzy case and several fuzzy extensions of DLs
can be found in the literature.  For a recent survey on the advances in the field of fuzzy DLs, we refer the
reader to~\cite{Lukasiewicz08a}. Besides the enrichment of DLs with fuzzy features, one of the challenges of the research in this community is the fact that different families of fuzzy operators (or fuzzy logics) lead to fuzzy DLs with different computational properties.  

Decidability of fuzzy DLs is often shown by adapting crisp DL tableau-based algorithms to the fuzzy DL case~\cite{Bobillo09c,Stoilos07b,Stoilos06,Straccia01,Straccia05d,Straccia07e}, or a reduction to classical DLs~\cite{Bobillo08d,Bobillo09,Bobillo08b,Bobillo11,Straccia04d}, or relying on some Mathematical Fuzzy Logic~\cite{HajekP98} based procedures~\cite{Cerami10,Cerdana10,HajekP05,HajekP06}.

However, recently there have been some unexpected surprises~\cite{Baader11,Baader11a,Bobillo11a}. Indeed, unlike the classical case, for the DL \alc~with GCIs
\emph{(i)} \cite{Bobillo11a} shows that it does not have the finite model property under  \L ukasiewicz Logic or Product Logic, illustrates that some algorithms are neither complete not correct, and shows some interesting conditions under which decidability is still guaranteed; and
\emph{(ii)} \cite{Baader11,Baader11a} show that knowledge base satisfiability is an undecidable problem for it under Product Logic.
Also worth mentioning is \cite{Borgwardt11a}, which illustrates the undecidability of knowledge base satisfiability if one replaces the truth set $[0,1]$ with  complete De Morgan lattices equipped with a t-norm operator.

In this paper, we complete  the analysis by showing that  knowledge base satisfiability is  an undecidable problem for the DL \alc~with GCIs under $[0,1]$-valued  \L ukasiewicz Logic as well. We prove our result following conceptually the methods devised in~\cite{Baader11,Baader11a,Borgwardt11a}.

We next introduce briefly our fuzzy DL, then we illustrate the undecidability result.

\section{The FDL \L-$\mathcal{ALC}$}

In this section we are going to introduce the general definitions of \L-$\mathcal{ALC}$ based on \L ukasiewicz $t$-norm.

\paragraph{Syntax.} Let ${\bf A}$ be a set of \emph{concept names}, ${\bf R}$ be a set of \emph{role names}. Concept names denote unary predicates, while role names denote binary predicates. The set of \L-$\mathcal{ALC}$ \emph{concepts} are built from concept names $A$ (also called atomic concepts) using connectives  and quantification constructs over roles $R$~\footnote{Each symbol may have super- and/or subscripts.} as described by the following syntactic rules:
\[
C\quad\rightarrow \quad \top\quad |\quad \bot \quad | \quad A\quad |\quad C_{1} \andc C_{2} \quad |\quad C_{1} \orc C_{2} \quad |\quad \notc C \quad |\quad \some R.C\quad |\quad  \all R.C \ .
\]

\nd An \emph{assertion} axiom is an expression of the form \fuzzyg{\cass{a}{C}}{n} (\emph{concept assertion}, $a$ is an instance of concept $C$ to degree at least $n$) or of the form $\fuzzyg{\rass{a_{1}}{1_{2}}{R}}{n}$ (\emph{role assertion}, $(a_{1}, a_{2})$ is an instance of role $R$ to degree at least $n$), where $a, a_{1}, a_{2}$ are individual names, $C$ is a concept, $R$ is a role name and $n \in (0,1]$ is a rational (a truth value). An \emph{ABox} $\A$  consists of a finite set of assertion axioms.

A  \emph{General Concept Inclusion} (GCI) axiom is of the form 
$\fuzzyg{C_{1} \impc C_{2}}{n}$
($C_{1}$ is a sub-concept of  $C_{2}$ to degree at least $n$), where $C_{i}$ is a concept and $n \in (0,1]$ is a rational. A \emph{concept hierarchy} $\T$, also called \emph{TBox}, is a finite set of GCIs.  In what follows we will use the following shorthands:
\begin{itemize}
\item $C_{1} \impc C_{2}$ for $\fuzzyg{C_{1} \impc C_{2}}{1}$ and $\cass{a}
{C}$ for $\fuzzyg{\cass{a}{C}}{1}$;

\item $C_{1} \defc C_{2}$ for the two axioms $C_{1} \impc C_{2}$ and $C_{2} \impc C_{1}$;

\item  $C_{1} \rimp C_{2}$ for $\neg C_{1} \orc C_{2}$;

\item  $C_{1} \requiv C_{2}$ for $(C_{1} \rimp C_{2}) \andc (C_{2} \rimp C_{1})$;

\item $\min\{C_{1}, C_{2}\}$ for $C_{1} \andc (C_{1} \rimp C_{2})$, and $\min\{C_{1}, \ldots, C_{n}\}$ for $\min\{\ldots \min\{C_{1}, C_{2}\}, \ldots \}$;

\item $\max\{C_{1}, C_{2}\}$ for $(C_{1} \rimp C_{2}) \rimp C_{2}$ and $\max\{C_{1}, \ldots, C_{n}\}$ for $\max\{\ldots \max\{C_{1}, C_{2}\}, \ldots\}$;


\item $n\cdot C$ for the $n$-ary disjunction $C\sqcup\ldots\sqcup C$;

\end{itemize}

\nd Finally, a \emph{knowledge base} $\KB = \tuple{\T, \A}$ consists of a TBox $\T$ and an ABox $\A$.

\paragraph{Semantics.}
From a semantics point of view, an axiom $\fuzzyg{\alpha}{n}$ constrains the truth degree of the expression $\alpha$ to be at least $n$. In the following, we use $\otimes, \oplus, \ominus$ and $\Rightarrow$ to denote \L ukasiewicz $t$-norm, $t$-conorm, negation function, and implication function, respectively~\cite{Klement00}. They are defined as operations in $[0,1]$ by means of the following functions:
\begin{eqnarray*}
a\otimes b & = & \max\{ 0, a+b-1\} \\ 
a\oplus b & = & \min\{ 1,a+b\} \\
\ominus a & = & 1-a \\
a \impf b & = &\min\{1,1-a+b\} \ ,
\end{eqnarray*}
\nd where $a$ and $b$ are arbitrary elements in $[0,1]$.
\nd As in the classical framework, the implication can be defined in terms of disjunction (whose semantics is the $t$-conorm) and negation in the usual way:
$a\impf b =\ominus a\oplus b$. Note also that for any implication defined from a continuous t-norm $\otimes$, it holds that:   
$x \Rightarrow y  = \max\{z \mid x \otimes z \leq y \}$, which is equivalent to the condition:
$y \geq x \otimes z$ iff $(x \Rightarrow y) \geq z$.

A \emph{fuzzy interpretation (or model)} is a pair $\I =
(\highi{\Delta}, \highi{\cdot})$ consisting of a nonempty (crisp)
set $\highi {\Delta}$ (the \emph{domain}) and of a \emph{fuzzy
interpretation function\/} $\highi{\cdot}$ that assigns:
\begin{enumerate}

    \item to each atomic concept $A$ a function
      $\highi{A}\colon\highi{\Delta} \rightarrow \unit$,

    \item to each role $R$ a function $\highi{R}\colon\highi{\Delta}
      \times \highi{\Delta} \rightarrow \unit$,

    \item to each individual $a$ an element $\highi{a} \in \highi{\Delta}$ such that $a^{\mathcal{I}} \neq b^{\mathcal{I}}$ if $a \neq b$ (\emph{Unique Name Assumption}, different individuals denote different objects of the domain).
\end{enumerate}

\begin{table*}
\[
\begin{array}{rcl}
\highi{\bottomc}(x) & = & 0 \\
\highi{\topc}(x) & = & 1 \\
\highi{(C \andc D)}(x) & = & \highi{C}(x) \andf \highi{D}(x)\\
\highi{(C \orc D)}(x) & = & \highi{C}(x) \orf \highi{D}(x) \\
\highi{(\notc C)}(x) & = & \notf \highi{C}(x)\\
\highi{(\all R.C)}(x) & = & \inf_{y \in \highi{\Delta}} \{ \highi{R}(x,y) \impf \highi{C}(y) \} \\
\highi{(\csome R.C)}(x) & = & \sup_{y \in \highi{\Delta}} \{ \highi{R}(x,y) \andf \highi{C}(y) \} \\
\end{array}
\]
\caption{Semantics for \L-\alc.} \label{DLsem}
\end{table*}

The fuzzy interpretation function is extended to 
complex concepts as specified in Table~\ref{DLsem} (where $x,y
\in \highi{\Delta}$ are elements of the domain). Hence, for every
complex concept $C$ we get a function $C^{\I}: \highi{\Delta}
\to [0,1]$.
The \emph{satisfiability of axioms} is then defined by the
following conditions:
\begin{enumerate}
  \item $\I$ satisfies an axiom $\fuzzyg{\cass{a}{C}}{\alpha}$ if    $C^{\I} (a^{\I}) \geq \alpha$,
  \item $\I$ satisfies an axiom $\fuzzyg{\rass{a}{b}{R}}{\alpha}$ if  $R^{\I} (a^{\I}, b^{\I}) \geq \alpha$,
  \item $\I$ satisfies an axiom $\fuzzyg{C \impc D}{\alpha}$ if $\highi{(C \impc D)} \geq \alpha$ where 
  \[
  \highi{(C \impc D)} =    \inf_{x \in \highi{\Delta}} \{ \highi{C}(x) \impf \highi{D}(x) \} \ .
  \]
\end{enumerate}

\nd It is interesting to point out that the satisfaction of a GCI of the
form $\fuzzyg{C \impc D}{1}$ is exactly the requirement that
$\forall x \in \Delta^{\I}, \highi{C}(x) \leq \highi{D}(x)$ (i.e.,
Zadeh's set inclusion); hence, in this particular case for the
satisfaction it only matters the partial order and not the exact
value of the implication $\Rightarrow$.

As it is expected we will say that a fuzzy interpretation $\I$
satisfies a KB $\KB$ in case that it satisfies all axioms in $\KB$.
And it is said that a fuzzy KB $\KB$ is \emph{satisfiable} iff there
exist a fuzzy interpretation $\I$ satisfying every axiom in $\KB$.

In this paper, we mainly focus on witnessed models. This notion (see~\cite{HajekP05}) corresponds to the restriction to the DL
language of the notion of witnessed model introduced, in the context
of the first-order language, by H\'ajek in \cite{HajekP07}. Specifically, a fuzzy
interpretation $\I$ is said to be \emph{witnessed} iff it holds that
for every complex concepts $C, D$, every role $R$, and every $x \in
\highi{\Delta}$ there is some
\begin{enumerate}
\item $y \in \highi{\Delta}$ such that
$(\some R.C)^{\I}(x) = R^{\I}(x,y) \otimes C^{\I}(y)$

\item $y \in \highi{\Delta}$ such that
$(\forall R.C)^{\I}(x) = R^{\I}(x,y) \Rightarrow C^{\I}(y)$

\end{enumerate}

\nd If $\I$ satisfies only condition 1.  then $\I$ is said to be \emph{weakly witnessed}. Note that for  \L ukasiewicz logic, condition 1. and 2. are equivalent, so $\I$ is weakly witnessed iff $I$ is witnessed. Thorough the paper we will rely on the notion of witnessed interpretation only, but keep in mind that the results apply, thus, to weakly witnessed interpretations as well. Note also that it is obvious that all finite fuzzy interpretations (this means that $\highi{\Delta}$ is a finite set) are indeed strongly witnessed but the opposite is not true. 

Sometimes (see, \eg~\cite{Baader11a}), the notion of witnessed interpretatations is strengthened to so-called \emph{strongly witnessed} interpretations by imposing that additionally that for every complex concepts $C, D$ and every $x \in \highi{\Delta}$ there is some

\begin{itemize}
\item $y \in \highi{\Delta}$ such that $(C \sqsubseteq D)^{\I} = C^{\I}(y) \Rightarrow D^{\I}(y)$
\end{itemize}

\nd has to hold. We do not deal with strongly witnessed interpretations here.

A fuzzy KB $\KB$ is said to be \emph{satisfiable} iff there exist a  fuzzy interpretation $\I$ satisfying every axiom in $\KB$.

\section{Undecidability of \L-\alc~with GCIs} \label{undec}

Our proof consists in a reduction of the \emph{reverse} of the \emph{Post Correspondence Problem} (PCP) and follows conceptually the one in~\cite{Baader11,Baader11a,Borgwardt11a}.
PCP is well-known to be undecidable~\cite{Post46}, so is the reverse PCP, as shown next.
\begin{definition}[PCP]
Let $v_{1},\ldots ,v_{p}$ and $w_{1},\ldots ,w_{p}$ be two finite lists of words over an alphabet $\Sigma = \{ 1,\ldots ,s\}$. The \emph{Post Correspondence Problem} (PCP) asks whether there is a non-empty sequence $i_{1},i_{2},\ldots ,i_{k}$, with $1\leq i_{j}\leq p$ such that $v_{i_{1}}v_{i_{2}}\ldots v_{i_{k}}=w_{i_{1}}w_{i_{2}}\ldots w_{i_{k}}$. Such a sequence, if it exists, is called a \emph{solution} of the problem instance.
\end{definition}

For the sake of our purpose, we will rely on a variant of the PCP, which we call \emph{Reverse} PCP (RPCP). Essentially, words are concatenated from  right to left rather than  from left to right.

\begin{definition}[RPCP]
Let $v_{1},\ldots ,v_{p}$ and $w_{1},\ldots ,w_{p}$ be two finite lists of words over an alphabet $\Sigma = \{ 1,\ldots ,s\}$. The \emph{Reverse Post Correspondence Problem} (RPCP) asks whether there is a non-empty sequence $i_{1},i_{2},\ldots ,i_{k}$, with $1\leq i_{j}\leq p$ such that $v_{i_{k}}v_{i_{k-1}}\ldots v_{i_{1}}=w_{i_{k}}w_{i_{k-1}}\ldots w_{i_{1}}$. Such a sequence, if it exists, is called a \emph{solution} of the problem instance.
\end{definition}

For a word $\mu=i_{1}i_{2}\ldots i_{k}\in\{ 1,\ldots ,p\}^{\ast}$ we will use $v_{\mu}$, $w_{\mu}$ to denote the words $v_{i_{k}} v_{i_{k-1}}\ldots  v_{i_{1}}$ and $w_{i_{k}}w_{i_{k-1}}\ldots w_{i_{1}}$. We denote the empty string as $\epsilon$ and define  $v_{\epsilon}$ is $\epsilon$.
The alphabet $\Sigma$ consists of the first $s$ positive integers. We can thus view every word in $\Sigma^{\ast}$ as a natural number represented in base $s+1$ in which 0 never occurs. Using this intuition, we will use the number 0 to encode the empty word.

Now we show that the reduction from PCP to RPCP is a very simple matter and it can be done through the transformation of the instance lists to the lists of their palindromes defined as follows:
let $\Sigma = \{ 1,\ldots ,s\}$ be an alphabet and $v=t_{1}t_{2}\ldots t_{|v|}$ a word over $\Sigma$, with $t_{i}\in\Sigma$, for $1\leq j\leq |v|$, then the \emph{palindrome} of $v$ is defined as $pal(v)=t_{|v|}t_{|v|-1}\ldots t_{1}$.

\begin{lemma}\label{lemma:pal}

Let $v_{1},\ldots ,v_{p}$ and $w_{1},\ldots ,w_{p}$ be two finite lists of words over an alphabet $\Sigma = \{ 1,\ldots ,s\}$. For every non-empty sequence $i_{1},i_{2},\ldots ,i_{k}$, with $1\leq i_{j}\leq p$ it holds that 
\begin{eqnarray*}
v_{i_{1}}v_{i_{2}}\ldots v_{i_{k}} & = & w_{i_{1}}w_{i_{2}}\ldots w_{i_{k}} \\
 & \mbox{iff} & \\
pal(v_{i_{k}}) pal(v_{i_{k-1}})\ldots pal(v_{i_{1}}) & = & pal(w_{i_{k}}) pal(w_{i_{k-1}})\ldots pal(w_{i_{1}}) \ .
\end{eqnarray*}
\end{lemma}
\begin{description}

\item[($Proof$)] First we prove by induction on $k$, that, for every sequence $v=v_{i_{1}}v_{i_{2}}\ldots v_{i_{k}}$ of words over $\Sigma$, it holds that $pal(v)=pal(v_{i_{k}})pal(v_{i_{k-1}})\ldots pal(v_{i_{1}})$.

\begin{itemize}

\item The case $k=1$ is straightforward.

\item Let $v=v_{i_{1}}v_{i_{2}}\ldots v_{i_{k}}$ and suppose, by inductive hypothesis, that $pal(v_{i_{1}}v_{i_{2}}\ldots v_{i_{k-1}})$ = $pal(v_{i_{k-1}})pal(v_{i_{k-2}})\ldots pal(v_{i_{1}})$. It follows that $pal(v)=pal(v_{i_{1}}v_{i_{2}}\ldots v_{i_{k-1}}, v_{i_{k}})=pal(v_{i_{k}})pal(v_{i_{k-1}})\ldots pal(v_{i_{1}})$.
\end{itemize}

Since the palindrome of a word is unique, we have that, if $v_{i_{1}}v_{i_{2}}\ldots v_{i_{k}} = w_{i_{1}}w_{i_{2}}\ldots w_{i_{k}}$, then $pal(v_{i_{1}}v_{i_{2}}\ldots v_{i_{k}}) = pal(w_{i_{1}}w_{i_{2}}\ldots w_{i_{k}})$ and, thus, $pal(v_{i_{k}})pal(v_{i_{k-1}})\ldots pal(v_{i_{1}}) = pal(w_{i_{k}})pal(w_{i_{k-1}})\ldots pal(w_{i_{1}})$. \qed
\end{description}

\begin{corollary}
The RPCP is undecidable.
\end{corollary}
\begin{description}
\item[($Proof$)] The proof is based on the reduction of PCPs to RCPs.
For every instance $\varphi=(v_{1},w_{1}), \ldots ,(v_{p},w_{p})$ of PCP, let $f$ be the function 
\[
f(\varphi)=(pal(v_{1}),pal(w_{1})), \ldots ,(pal(v_{p}),pal(w_{p})) \ .
\]
\nd Clearly $f$ is a computable function. Moreover,
%
%
%
$\varphi\in PCP$ if and only if there exists a non-empty sequence $i_{1},i_{2},\ldots ,i_{k}$, with $1\leq i_{j}\leq p$ such that $v_{i_{1}}v_{i_{2}}\ldots v_{i_{k}}=w_{i_{1}}w_{i_{2}}\ldots w_{i_{k}}$, that is, by Lemma \ref{lemma:pal}, 
\[
pal(v_{i_{k}})pal(v_{i_{k-1}})\ldots pal(v_{i_{1}}) = pal(w_{i_{k}})pal(w_{i_{k-1}})\ldots pal(w_{i_{1}})\]
\ie~$f(\varphi)\in RPCP$. Therefore, $\varphi\in PCP$ if and only if $f(\varphi)\in RPCP$.\qed



%
\end{description}


\paragraph{Undecidability of general KB satisfiability.} We show the undecidability by a reduction of RPCPs to KB satisfiability problems. Specifically, given an instance $\varphi$ of RPCP, we will construct a Knowledge Base $\mathcal{O}_{\varphi}$ that is satisfiable iff $\varphi$ has no solution. 

In order to do this we will encode words $v$ from the alphabet $\Sigma$ as rational numbers $0.v$ in $[0,1]$ in base $s+1$; the empty word will be encoded by the number 0.

So, let us define  the TBox
\begin{eqnarray*}
\mathcal{T} & :=  \{ &
V\equiv V_{1}\sqcup V_{2}, W \defc W_{1}\sqcup W_{2} \ \ \ \} 
\end{eqnarray*}

and for $1 \leq i \leq p$ the TBoxes


\begin{eqnarray*}
\mathcal{T}^{i}_{\varphi} & :=  \{ &
\top\sqsubseteq\exists R_{i}.\top,  \\\\
%
%
& &  V \impc (s+1)^{|v_{i}|}\cdot\forall R_{i}.V_{1}, \\
 & &   (s+1)^{|v_{i}|}\cdot\exists R_{i}.V_{1} \impc V, \\
&&  W \impc (s+1)^{|w_{i}|}\cdot\forall R_{i}.W_{1} ,  \\
&&  (s+1)^{|w_{i}|}\cdot\exists R_{i}.W_{1} \impc W \\\\
 & & \fuzzyg{\top\sqsubseteq\forall R_{i}.V_{2}}{0.v_{i}} , \\
&& \fuzzyg{\top\sqsubseteq\forall R_{i}.\neg V_{2}}{1-0.v_{i}} , \\
 & & \fuzzyg{\top\sqsubseteq\forall R_{i}.W_{2}}{0.w_{i}} , \\
&&\fuzzyg{\top\sqsubseteq\forall R_{i}.\neg W_{2}}{1-0.w_{i}} , \\\\
%
 %
 & &  A\impc (s+1)^{\max\{|v_{i}|,|w_{i}|\}}\cdot\forall R_{i}.A \\
 & &  (s+1)^{\max\{|v_{i}|,|w_{i}|\}}\cdot \exists R_{i}.A \impc A \ \ \ \} \ .
\end{eqnarray*}

Now, let 
\[
\mathcal{T}_{\varphi}=\mathcal{T} \cup \bigcup_{i=1}^{p}\mathcal{T}^{i}_{\varphi} \ .
\]
Further we define the ABox $\mathcal{A}$ as follows:
\begin{eqnarray*}
\mathcal{A} & := & \{ a:\neg V, a:\neg W, 
  \fuzzyg{a:A}{0.01} , \fuzzyg{a:\neg A}{0.99}\} \ .
\end{eqnarray*}
Finally, we define 
\[
\mathcal{O}_{\varphi}:=\tuple{\mathcal{T}_{\varphi},\mathcal{A}} \ .
\] 
We now define the interpretation 
\[
\mathcal{I}_{\varphi}:=(\Delta^{\mathcal{I}_{\varphi}},\cdot^{\mathcal{I}_{\varphi}})
\]
as follows:

\begin{itemize}

\item $\Delta^{\mathcal{I}_{\varphi}}=\{ 1,\ldots ,p\}^{*}$

\item $a^{\mathcal{I}_{\varphi}}=\epsilon$

\item $V^{\mathcal{I}_{\varphi}}(\epsilon) = W^{\mathcal{I}_{\varphi}}(\epsilon)=0$, $A^{\mathcal{I}_{\varphi}}(\epsilon) = 0.01$,
and for $1\leq i \leq 2$, $V_{i}^{\mathcal{I}_{\varphi}}(\epsilon)= W_{i}^{\mathcal{I}_{\varphi}}(\epsilon)=0$


\item  for all $\mu,\mu'\in\Delta^{\mathcal{I}_{\varphi}}$ and $1\leq i\leq p$
\[
R_{i}^{\mathcal{I}_{\varphi}}(\mu ,\mu')=\begin{cases}
1, & \text{if }\mu'=\mu i \\
0, & \text{otherwise}
\end{cases}\]

\item for every $\mu\in\Delta^{\mathcal{I}_{\varphi}}$, where $\mu = i_{1}i_{2} \ldots i_{k} \neq \epsilon$

\begin{itemize}

\item $V^{\mathcal{I}_{\varphi}}(\mu)=0.v_{\mu}$, $W^{\mathcal{I}_{\varphi}}(\mu)=0.w_{\mu}$


\item $A^{\mathcal{I}_{\varphi}}(\mu) = 0.01\cdot (s+1)^{-\sum_{j \in \{i_{1},i_{2}, \ldots, i_{k} \}}\max\{ |v_{j}|,|w_{j}|\}}$

\item $V_{1}^{\mathcal{I}_{\varphi}}(\mu)=0.v_{\bar{\mu}}\cdot (s+1)^{-|v_{i_{k}}|}$,
$W_{1}^{\mathcal{I}_{\varphi}}(\mu)=0.w_{\bar{\mu}}\cdot (s+1)^{-|w_{i_{k}}|}$, where
$\bar{\mu} = i_{1}i_{2} \ldots i_{k-1}$ (last index $ i_{k}$ is dropped from $\mu$, and we assume that $0.\epsilon$ is $0$),

\item $V_{2}^{\mathcal{I}_{\varphi}}(\mu)=0.v_{i_{k}}$, $W_{2}^{\mathcal{I}_{\varphi}}(\mu)=0.w_{i_{k}}$.

%
%
%
%
\end{itemize}
\end{itemize}

It is easy to see that $\mathcal{I}_{\varphi}$ is a witnessed model of $\mathcal{O}_{\varphi}$ (note that \eg~$(\forall R_{i}.V_{1})^{\mathcal{I}_{\varphi}}(\mu) = V_{1}^{\mathcal{I}_{\varphi}}(\mu i)$.~\footnote{However, $\mathcal{I}_{\varphi}$ is not a strongly witnessed model of $\mathcal{O}_{\varphi}$.}

Moreover, as in \cite{Baader11} it is possible to prove that, for every  witnessed model $\mathcal{I}$ of $\mathcal{O}_{\varphi}$, there is a mapping $g$ from $\mathcal{I}_{\varphi}$ to $\mathcal{I}$.

\begin{lemma}\label{lemma:homomorphism}

Let $\mathcal{I}$ be a  witnessed model of $\mathcal{O}_{\varphi}$. Then there exists a function $g :  \Delta^{\mathcal{I}_{\varphi}}\to\Delta^{\mathcal{I}}$ such that, for every $\mu\in\Delta^{\mathcal{I}_{\varphi}}$, $C^{\mathcal{I}_{\varphi}}(\mu)= C^{\mathcal{I}}(g(\mu))$  holds for every concept name $C$ and $R_{i}^{\mathcal{I}_{\varphi}}(\mu ,\mu i) = R_{i}^{\mathcal{I}}(g(\mu) ,g(\mu i))$ holds for every $i$, with $1\leq i\leq p$.

\end{lemma}

\begin{description}

\item[($Proof$)] Let $\mathcal{I}$ be a  witnessed model of $\mathcal{O}_{\varphi}$. We will build the function $g$ inductively on the length of $\mu$.

\begin{itemize}

\item[($\epsilon$)] Since $\mathcal{I}$ is a model of $\mathcal{O}_{\varphi}$, then there is an element $\delta\in\Delta^{\mathcal{I}}$ such that $a^{\mathcal{I}}=\delta$. Since $\mathcal{I}$ is a model of $\mathcal{A}_{\varphi}$, setting $g(\epsilon)=\delta$, we have that $V^{\mathcal{I}_{\varphi}}(\epsilon)=0= V^{\mathcal{I}}(g(\epsilon))$ and the same holds for concept $W$. Moreover, since $\mathcal{I}$ is a model of $\mathcal{T}_{\varphi}$, we have that $V^{\mathcal{I}}(\delta)=(V_{1}\sqcup V_{2})^{\mathcal{I}}(\delta)$ and, therefore $V_{1}^{\mathcal{I}_{\varphi}}(\epsilon)=0= V_{1}^{\mathcal{I}}(g(\epsilon))$ and the same holds for $V_{2}$, $W_{1}$ and $W_{2}$. On the other hand, we have that $A^{\mathcal{I}_{\varphi}}(\epsilon)=0.01= A^{\mathcal{I}}(g(\epsilon))$, as well. 
So, $g(\epsilon)=\delta$ satisfies the condition of the lemma.


\item[($\mu i$)] Let now $\mu$ be such that $g(\mu)$ has already been defined. Now, since $\mathcal{I}$ is a  witnessed model and satisfies axiom $\top\sqsubseteq\exists R_{i}.\top$, then for all $i$, with $1\leq i\leq p$, there exists a $\gamma\in\Delta^{\mathcal{I}}$ such that $R_{i}^{\mathcal{I}}(g(\mu),\gamma)=1$. So, setting 
$g(\mu i) = \gamma$ we get $1 = R_{i}^{\mathcal{I}_{\varphi}}(\mu , \mu i) =  R_{i}^{\mathcal{I}}(g(\mu),g(\mu i))$. Furthermore, by inductive hypothesis, we can assume that $V^{\mathcal{I}}(g(\mu))=0.v_{\mu}$ and $W^{\mathcal{I}}(g(\mu))=0.w_{\mu}$. 


Since $\mathcal{I}$ satisfies axiom $ V \impc (s+1)^{|v_{i}|}\cdot \forall R_{i}.V_{1}$, then
$0.v_{\mu} = V^{\mathcal{I}}(g(\mu)) \leq (s+1)^{|v_{i}|}\cdot  (\forall R_{i}.V_{1})^{\mathcal{I}}(g(\mu)) =(s+1)^{|v_{i}|}\cdot\inf_{\gamma\in\Delta^{\mathcal{I}}}\{ R_{i}^{\mathcal{I}}(g(\mu),\gamma)\Rightarrow V^{\mathcal{I}}_{1}(\gamma)\}\leq (s+1)^{|v_{i}|}\cdot R_{i}^{\mathcal{I}}(g(\mu),\mu i)\Rightarrow V^{\mathcal{I}}_{1}(\mu i) = (s+1)^{|v_{i}|}\cdot  V^{\mathcal{I}}_{1}(g(\mu i))$.

Since $\mathcal{I}$ satisfies axiom $(s+1)^{|v_{i}|}\cdot \exists R_{i}.V_{1} \impc  V$, then
$0.v_{\mu} = V^{\mathcal{I}}(g(\mu)) \geq (s+1)^{|v_{i}|}\cdot  (\exists R_{i}.V_{1})^{\mathcal{I}}(g(\mu)) =(s+1)^{|v_{i}|}\cdot\sup_{\gamma\in\Delta^{\mathcal{I}}}\{ R_{i}^{\mathcal{I}}(g(\mu),\gamma)\otimes V^{\mathcal{I}}_{1}(\gamma)\}\geq (s+1)^{|v_{i}|}\cdot R_{i}^{\mathcal{I}}(g(\mu),\mu i)\otimes V^{\mathcal{I}}_{1}(\mu i) = (s+1)^{|v_{i}|}\cdot  V^{\mathcal{I}}_{1}(g(\mu i))$. 
Therefore, $(s+1)^{|v_{i}|}\cdot  V^{\mathcal{I}}_{1}(g(\mu i))=0.v_{\mu}$ and
$V^{\mathcal{I}}_{1}(g(\mu i))=0.v_{\mu}\cdot (s+1)^{-|v_{i}|} = V^{\mathcal{I}_{\varphi}}_{1}(\mu i)$. 


Similarly, it can be shown that $W^{\mathcal{I}}_{1}(g(\mu i))=0.w_{\mu}\cdot (s+1)^{-|w_{i}|} = W^{\mathcal{I}_{\varphi}}_{1}(\mu i)$.

Since $\mathcal{I}$ satisfies axioms $\fuzzyg{\top\sqsubseteq\forall R_{i}.V_{2}}{0.v_{i}}$ and $\fuzzyg{\top\sqsubseteq\forall R_{i}.\neg V_{2}}{1-0.v_{i}}$,  it follows that
$(\forall R_{i}.V_{2})^{\mathcal{I}}(g(\mu)) \geq 0.v_{i}$ and 
$(\forall R_{i}. \neg V_{2})^{\mathcal{I}}(g(\mu)) \geq 1- 0.v_{i}$. Therefore, for
$R_{i}^{\mathcal{I}}(g(\mu),g(\mu i)) = 1$ we have
$V^{\mathcal{I}}_{2}(g(\mu i)) =  0.v_{i} =  V^{\mathcal{I}_{\varphi}}_{2}(\mu i)$.
Similarly, it can be shown that $W^{\mathcal{I}_{\varphi}}_{2}(\mu i) = 0.w_{i} = W^{\mathcal{I}}_{2}(g(\mu i))$.

Now, since $\mathcal{I}$ satisfies axiom $ V\defc V_{1}\sqcup V_{2}$, then,  
$V^{\mathcal{I}}(g(\mu i)) = V_{1}^{\mathcal{I}}(g(\mu i)) + V_{2}^{\mathcal{I}}(g(\mu i)) = 
0.v_{\mu}\cdot (s+1)^{-|v_{i}|} + 0.v_{i} = 0.v_{i}v_{\mu}=V^{\mathcal{I}_{\varphi}}(\mu i)$. 

Finally, by inductive hypothesis,  assume that $A^{\mathcal{I}}(g(\mu))=A^{\mathcal{I}_{\varphi}}(\mu) = 0.01\cdot (s+1)^{-\sum_{j \in \{i_{1},i_{2}, \ldots, i_{k} \}}\max\{ |v_{j}|,|w_{j}|\}}$, where $\mu = i_{1}i_{2} \ldots i_{k}$. 
%


Since $\mathcal{I}$ satisfies axioms $A\impc (s+1)^{\max\{|v_{i}|,|w_{i}|\}}\cdot \forall R_{i}.A$, we have that 
\[
A^{\mathcal{I}}(g(\mu )) \leq (s+1)^{\max\{|v_{i}|,|w_{i}|\}}\cdot (\forall R_{i}.A)^{\mathcal{I}}(g(\mu)) \leq (s+1)^{\max\{|v_{i}|,|w_{i}|\}}\cdot A^{\mathcal{I}}(g(\mu i)) \ . 
\]
Likewise, since $\mathcal{I}$ satisfies axioms $ (s+1)^{\max\{|v_{i}|,|w_{i}|\}}\cdot \exists R_{i}.A \impc A$, we have that 
\[
A^{\mathcal{I}}(g(\mu )) \geq (s+1)^{\max\{|v_{i}|,|w_{i}|\}}\cdot (\exists R_{i}.A)^{\mathcal{I}}(g(\mu)) \geq (s+1)^{\max\{|v_{i}|,|w_{i}|\}}\cdot A^{\mathcal{I}}(g(\mu i)) \ 
\]
and, thus, 
\[
A^{\mathcal{I}}(g(\mu )) =  (s+1)^{\max\{|v_{i}|,|w_{i}|\}}\cdot A^{\mathcal{I}}(g(\mu i)) \ . 
\]


Therefore,
\begin{eqnarray*}
A^{\mathcal{I}}(g(\mu i)) & = & (s+1)^{-\max\{|v_{i}|,|w_{i}|\}}\cdot A^{\mathcal{I}}(g(\mu )) \\
& = & (s+1)^{-\max\{|v_{i}|,|w_{i}|\}}\cdot A^{\mathcal{I}_{\varphi}}(\mu) \\
& = & (s+1)^{-\max\{|v_{i}|,|w_{i}|\}}\cdot 0.01\cdot (s+1)^{-\sum_{j \in \{i_{1},i_{2}, \ldots, i_{k} \}}\max\{ |v_{j}|,|w_{j}|\}}\\
& = & 0.01\cdot  (s+1)^{- (\max\{|v_{i}|,|w_{i}|\} +\sum_{j \in \{i_{1},i_{2}, \ldots, i_{k} \}}\max\{ |v_{j}|,|w_{j}|\})} \\
&= & 0.01\cdot  (s+1)^{-  \sum_{j \in \{i_{1},i_{2}, \ldots, i_{k},i \}}\max\{ |v_{j}|,|w_{j}|\}} \\
& = & A^{\mathcal{I}_{\varphi}}(\mu i) \ ,
\end{eqnarray*}

which completes the proof.\qed

\end{itemize}

\end{description}

From the last Lemma it follows that if the RPCP instance $\varphi$ has a solution $\mu$, for some $\mu\in\{ 1,\ldots ,p\}^{+}$, then $v_{\mu} = w_{\mu}$ and, thus, $0.v_{\mu} = 0.w_{\mu}$. Therefore, every  witnessed model $\mathcal{I}$ of $\mathcal{O}_{\varphi}$ contains an element $\delta=g(\mu)$ such that $V^{\mathcal{I}}(\delta)= V^{\mathcal{I}_{\phi}}(\mu) = 0.v_{\mu} = 0.w_{\mu} =  W^{\mathcal{I}_{\phi}}(\mu)= W^{\mathcal{I}}(\delta)$. Conversely, from the definition of $\mathcal{I}_{\varphi}$, if $\varphi$ has no solution, then there is no $\mu$ such that $0.v_{\mu} = 0.w_{\mu}$, \ie~there is no $\mu$  such that $V^{\mathcal{I}_{\phi}}(\mu) = W^{\mathcal{I}_{\phi}}(\mu)$. 


However, as $\mathcal{O}_{\varphi}$ is always satisfiable, it does not yet help us to decide the RPCP. We next extend $\mathcal{O}_{\varphi}$ to $\mathcal{O}'_{\varphi}$ in such a way that an instance $\varphi$ of the RPCP has a solution iff the ontology $\mathcal{O}_{\varphi}'$ is not  witnessed satisfiable and, thus, establish that the KB satisfiability problem is undecidable.
To this end, consider
\[
\mathcal{O}_{\varphi}':=\tuple{\mathcal{T}_{\varphi}',\mathcal{A}} \ ,
\]
where
\[
\mathcal{T}_{\varphi}':=\mathcal{T}_{\varphi}\cup\bigcup_{1\leq i\leq p}\{\top\sqsubseteq\forall R_{i}.(\neg(V\requiv W)\sqcup\neg A)\} \ .
\]


The intuition here is the following. If there is a solution for RPCP then, by the observation before, there is a point $\delta$ in which the value of $V$ and $W$ coincide under $\mathcal{I}$. That is, the value of $\neg(V\requiv W)$ is $0$ and, thus, the one of $\neg(V\requiv W)\sqcup\neg A)$ is less than $1$. So, $\mathcal{I}$ cannot satisfy the new GCI in $\mathcal{T}_{\varphi}'$ and, thus, $\mathcal{O}_{\varphi}'$ is not satisfiable. On the other hand, if there is no solution to the RPCP then in $\mathcal{I}_{\varphi}$ there is no point in which $V$ and $W$ coincide and, thus, $\neg(V\requiv W) > 0$. However, we will show that the value of $\neg(V\requiv W)$ in all  points is strictly greater than $A$ and, as $A \orc \neg A$ is $1$, so also  $\neg(V\requiv W)\sqcup\neg A$ will be $1$ in any point. Hence, $\mathcal{I}_{\phi}$ is a model of the aditional axiom in $\mathcal{T}_{\varphi}'$, \ie~$\mathcal{O}_{\varphi}'$ is satisfiable. 

\begin{theorem}

The instance $\varphi$ of the RPCP has a solution iff the ontology $\mathcal{O}_{\varphi}'$ is not  witnessed satisfiable.

\end{theorem}

\begin{description}

\item[($Proof$)] Assume first that $\varphi$ has a solution $\mu=i_{1}\ldots i_{k}$ and let $\mathcal{I}$ be a  witnessed model of $\mathcal{O}_{\varphi}$. Let $\bar{\mu} = i_{1}i_{2} \ldots i_{k-1}$ (last index $ i_{k}$ is dropped from $\mu$). Then by Lemma \ref{lemma:homomorphism} it follows that there are nodes $\delta ,\delta'\in\Delta^{\mathcal{I}}$ such that  $\delta=g(\mu)$,  $\delta'=g(\bar{\mu})$, with $V^{\mathcal{I}}(\delta)=V^{\mathcal{I}_{\varphi}}(\mu)=W^{\mathcal{I}_{\varphi}}(\mu)=W^{\mathcal{I}}(\delta)$ and $R^{\mathcal{I}}_{i_{k}}(\delta',\delta)=1$. Then $(V\requiv W)^{\mathcal{I}}(\delta)=1$. Since $(\neg A)^{\mathcal{I}}(\delta)<1$, then $(\neg(V\requiv W)\sqcup\neg A)^{\mathcal{I}}(\delta)<1$. Hence there is $i$, with $1\leq i\leq p$, such that $(\forall R_{i}.(\neg(V\requiv W)\sqcup\neg A))^{\mathcal{I}}(\delta')<1$. So, axiom $\top\sqsubseteq\forall R_{i}.(\neg(V\requiv W)\sqcup\neg A)$ is not satisfied and, therefore, $\mathcal{O}_{\varphi}$ is not satisfiable.

For the converse, assume that $\varphi$ has no solution. On the one hand we know that $\mathcal{I}_{\varphi}$ is a model of $\mathcal{O}_{\varphi}$. On the other hand, since $\varphi$ has no solution, then there is no $\mu=i_{1}\ldots i_{k}$ such that $v_{\mu}=w_{\mu}$ (\ie $0.v_{\mu} = 0.w_{\mu}$) and, therefore, there is no $\mu\in\Delta^{\mathcal{I}_{\varphi}}$ such that $V^{\mathcal{I}_{\varphi}}(\mu)=W^{\mathcal{I}_{\varphi}}(\mu)$. Consider $\mu\in\Delta^{\mathcal{I}_{\varphi}}$ and $i$, with $1\leq i\leq p$ and assume, without loss of generality, that $V^{\mathcal{I}_{\varphi}}(\mu i)<W^{\mathcal{I}_{\varphi}}(\mu i)$. Then 
\begin{eqnarray*}
(V\requiv W)^{\mathcal{I}_{\varphi}}(\mu i) & = & (V^{\mathcal{I}_{\varphi}}(\mu i) \Rightarrow W^{\mathcal{I}_{\varphi}}(\mu i)) \otimes (W^{\mathcal{I}_{\varphi}}(\mu i) \Rightarrow V^{\mathcal{I}_{\varphi}}(\mu i)) \\
& = & 1 \otimes (W^{\mathcal{I}_{\varphi}}(\mu i) \Rightarrow V^{\mathcal{I}_{\varphi}}(\mu i))  \\
& = & W^{\mathcal{I}_{\varphi}}(\mu i) \Rightarrow V^{\mathcal{I}_{\varphi}}(\mu i) \\
& = & 1 - W^{\mathcal{I}_{\varphi}}(\mu i) + V^{\mathcal{I}_{\varphi}}(\mu i) \\
& = & 1 - (W^{\mathcal{I}_{\varphi}}(\mu i) - V^{\mathcal{I}_{\varphi}}(\mu i)) \\
& = & 1-(0.w_{\mu i}-0.v_{\mu i}) \\
& \leq & 1-0.01\cdot (s+1)^{-\max\{|v_{\mu i}|,|w_{\mu i}|\}}\\
& \leq & 1- 0.01\cdot  (s+1)^{-  \sum_{j \in \{i_{1},i_{2}, \ldots, i_{k},i \}}\max\{ |v_{j}|,|w_{j}|\}} \\
& = & (\neg A)^{\mathcal{I}_{\varphi}}(\mu i) \ .
\end{eqnarray*}


Therefore, $(\neg (V\requiv W))^{\mathcal{I}_{\varphi}}(\mu i) \geq  A^{\mathcal{I}_{\varphi}}(\mu i)$. As
$A^{\mathcal{I}_{\varphi}}(\mu i) \oplus (\neg A)^{\mathcal{I}_{\varphi}}(\mu i) = 1$, it follows that for every $\mu\in\Delta^{\mathcal{I}_{\varphi}}$ and $i$, with $1\leq i\leq p$, it holds that $(\forall R_{i}.(\neg(V\requiv W)\sqcup\neg A))^{\mathcal{I}_{\varphi}}(\mu) =1$ and, therefore, $\mathcal{I}_{\varphi}$ is a  witnessed model of $\mathcal{O}_{\varphi}'$. \qed
\end{description}


\begin{thebibliography}{10}

\bibitem{Baader03a}
Franz Baader, Diego Calvanese, Deborah McGuinness, Daniele Nardi, and Peter~F.
  Patel-Schneider, editors.
\newblock {\em The Description Logic Handbook: Theory, Implementation, and
  Applications}.
\newblock Cambridge University Press, 2003.

\bibitem{Baader11}
Franz Baader and Rafael Pe{\~n}aloza.
\newblock Are fuzzy description logics with general concept inclusion axioms
  decidable?
\newblock In {\em Proceedings of 2011 IEEE International Conference on Fuzzy
  Systems ({Fuzz-IEEE 2011})}. IEEE Press, 2011.

\bibitem{Baader11a}
Franz Baader and Rafael Pe{\~n}aloza.
\newblock {GCI}s make reasoning in fuzzy {DLs} with the product t-norm
  undecidable.
\newblock In {\em Proceedings of the 24th International Workshop on Description
  Logics (DL-11)}. CEUR Electronic Workshop Proceedings, 2011.

\bibitem{Bobillo11a}
Fernando Bobillo, F\'{e}lix Bou, and Umberto Straccia.
\newblock On the failure of the finite model property in some fuzzy description
  logics.
\newblock {\em Fuzzy Sets and Systems}, 172(1):1--12, 2011.

\bibitem{Bobillo08d}
Fernando Bobillo, Miguel Delgado, and Juan G\'{o}mez-Romero.
\newblock A crisp representation for fuzzy {$\mathcal{SHOIN}$} with fuzzy
  nominals and general concept inclusions.
\newblock In {\em Uncertainty Reasoning for the Semantic Web I}, volume 5327 of
  {\em Lecture Notes in Computer Science}, pages 174--188. Springer Verlag,
  2008.

\bibitem{Bobillo09}
Fernando Bobillo, Miguel Delgado, Juan G\'{o}mez-Romero, and Umberto Straccia.
\newblock Fuzzy description logics under g{\"o}del semantics.
\newblock {\em International Journal of Approximate Reasoning}, 50(3):494--514,
  2009.

\bibitem{Bobillo08b}
Fernando Bobillo and Umberto Straccia.
\newblock On qualified cardinality restrictions in fuzzy description logics
  under {{\L}ukasiewicz} semantics.
\newblock In Luis Magdalena, Manuel Ojeda-Aciego, and Jos\'{e}~Luis Verdegay,
  editors, {\em Proceedings of the 12th International Conference of Information
  Processing and Management of Uncertainty in Knowledge-Based Systems (IPMU
  2008)}, pages 1008--1015, June 2008.

\bibitem{Bobillo09c}
Fernando Bobillo and Umberto Straccia.
\newblock Fuzzy description logics with general t-norms and datatypes.
\newblock {\em Fuzzy Sets and Systems}, 160(23):3382---3402, 2009.

\bibitem{Bobillo11}
Fernando Bobillo and Umberto Straccia.
\newblock Reasoning with the finitely many-valued lukasiewicz fuzzy description
  logic {SROIQ}.
\newblock {\em Information Sciences}, pages xxx--xxx, 2011.

\bibitem{Borgwardt11a}
Stefan Borgwardt and Rafael Pe{\~n}aloza.
\newblock Fuzzy ontologies over lattices with t-norms.
\newblock In {\em Proceedings of the 24th International Workshop on Description
  Logics (DL-11)}. CEUR Electronic Workshop Proceedings, 2011.
\newblock To appear.

\bibitem{Cerami10}
Marco Cerami, Francesc Esteva, and F\`elix Bou.
\newblock Decidability of a description logic over infinite-valued product
  logic.
\newblock In {\em Proceedings of the Twelfth International Conference on
  Principles of Knowledge Representation and Reasoning (KR-10)}. AAAI Press,
  2010.

\bibitem{Cerdana10}
\`{A}ngel Garc\i\'{}a-Cerda\`{o}a, Eva Armengol, and Francesc Esteva.
\newblock Fuzzy description logics and t-norm based fuzzy logics.
\newblock {\em Int. J. Approx. Reasoning}, 51:632--655, July 2010.

\bibitem{HajekP98}
Petr H{\'a}jek.
\newblock {\em Metamathematics of Fuzzy Logic}.
\newblock Kluwer, 1998.

\bibitem{HajekP05}
Petr H{\'a}jek.
\newblock Making fuzzy description logics more expressive.
\newblock {\em Fuzzy Sets and Systems}, 154(1):1--15, 2005.

\bibitem{HajekP06}
Petr H{\'a}jek.
\newblock What does mathematical fuzzy logic offer to description logic?
\newblock In Elie Sanchez, editor, {\em Fuzzy Logic and the Semantic Web},
  Capturing Intelligence, chapter~5, pages 91--100. Elsevier, 2006.

\bibitem{HajekP07}
Petr H{\'a}jek.
\newblock On witnessed models in fuzzy logic.
\newblock {\em Mathematical Logic Quarterly}, 53(1):66--77, 2007.

\bibitem{Klement00}
Erich~Peter Klement, Radko Mesiar, and Endre Pap.
\newblock {\em Triangular Norms}.
\newblock Trends in Logic - Studia Logica Library. Kluwer Academic Publishers,
  2000.

\bibitem{Lukasiewicz08a}
Thomas Lukasiewicz and Umberto Straccia.
\newblock Managing uncertainty and vagueness in description logics for the
  semantic web.
\newblock {\em Journal of Web Semantics}, 6:291--308, 2008.

\bibitem{OWL2}
{\mbox{{OWL 2 Web Ontology Language} Document Overview}}.
\newblock {\em \url{http://www.w3.org/TR/2009/REC-owl2-overview-20091027/}}.
\newblock {W3C}, 2009.

\bibitem{Post46}
Emil~L. Post.
\newblock A variant of a recursively unsolvable problem.
\newblock {\em Bulletin of The American Mathematical Society}, 52:264--269,
  1946.

\bibitem{Stoilos07b}
Giorgos Stoilos, Giorgos~B. Stamou, Jeff~Z. Pan, Vassilis Tzouvaras, and Ian
  Horrocks.
\newblock Reasoning with very expressive fuzzy description logics.
\newblock {\em Journal of Artificial Intelligence Research}, 30:273--320, 2007.

\bibitem{Stoilos06}
Giorgos Stoilos, Umberto Straccia, Giorgos Stamou, and Jeff Pan.
\newblock General concept inclusions in fuzzy description logics.
\newblock In {\em Proceedings of the 17th Eureopean Conference on Artificial
  Intelligence (ECAI-06)}, pages 457--461. IOS Press, 2006.

\bibitem{Straccia01}
Umberto Straccia.
\newblock Reasoning within fuzzy description logics.
\newblock {\em Journal of Artificial Intelligence Research}, 14:137--166, 2001.

\bibitem{Straccia04d}
Umberto Straccia.
\newblock Transforming fuzzy description logics into classical description
  logics.
\newblock In {\em Proceedings of the 9th European Conference on Logics in
  Artificial Intelligence (JELIA-04)}, number 3229 in Lecture Notes in Computer
  Science, pages 385--399, Lisbon, Portugal, 2004. Springer Verlag.

\bibitem{Straccia05d}
Umberto Straccia.
\newblock Description logics with fuzzy concrete domains.
\newblock In Fahiem Bachus and Tommi Jaakkola, editors, {\em 21st Conference on
  Uncertainty in Artificial Intelligence (UAI-05)}, pages 559--567, Edinburgh,
  Scotland, 2005. AUAI Press.

\bibitem{Straccia07e}
Umberto Straccia and Fernando Bobillo.
\newblock Mixed integer programming, general concept inclusions and fuzzy
  description logics.
\newblock {\em Mathware \& Soft Computing}, 14(3):247--259, 2007.

\end{thebibliography}

\end{document}